# Topological insights from state manipulation in a classical elastic system

Kazi T. Mahmood ; M. Arif Hasan

# Topological insights from state manipulation in a classical elastic system

Kazi T. Mahmood[a] and M. Arif Hasan[b]

## AFFILIATIONS

Department of Mechanical Engineering, Wayne State University, Detroit, Michigan 48202, USA

[a]kazi.tahsin.mahmood@wayne.edu
[b]**Author to whom correspondence should be addressed:** hasan.arif@wayne.edu

## ABSTRACT

The exploration of the Berry phase in classical mechanics has opened new frontiers in understanding the dynamics of physical systems, analogous to quantum mechanics. Here, we show controlled accumulation of the Berry phase in a two-level elastic bit, which is a classical counterpart to qubits, achieved by manipulating coupled granules with external drivers. Employing the Bloch sphere representation, the paper demonstrates the manipulation of elastic bit states and the realization of quantum-analog logic gates. A key achievement is the calculation of the Berry phase for various system states, revealing insights into the system's topological nature. Unique to this study is the use of external parameters to explore topological transitions, contrasting with traditional approaches focusing on internal system modifications. By linking the classical and quantum worlds through the Berry phase of an elastic bit, this work extends the potential applications of topological concepts in designing new materials and computational models.

## I. INTRODUCTION

The study of geometric and Berry phases has profoundly enriched the realms of classical and quantum mechanics. These phases, arising during cyclic adiabatic processes, are far from mere mathematical constructs; they provide deep insights into the essence of physical systems. One remarkable aspect of the geometric phase is its topological nature. It is insensitive to small perturbations or deformations of the path but depends only on global properties such as winding numbers or topology. This robustness makes it particularly relevant for quantum computation and information processing applications. In quantum computation, geometric phases can be utilized for implementing quantum gates and performing fault-tolerant operations on qubits. By carefully designing paths in parameter space, one can manipulate qubits without being affected by certain types of noise or decoherence processes.[1–3] Topological insulators are another area where geometric phases are crucial.[4] These materials exhibit unique electronic properties due to their nontrivial topology. Geometric phases have also been observed in various other systems, from neutrons' behavior to twisted anisotropic materials' properties.

The Berry phase, a topological interpretation of the geometric phase, extends beyond quantum physics and mathematics, influencing classical mechanics as well.[5–8] In classical mechanics, topology relates to how system components are arranged and interlinked, which defines the system's structure and functionality. One example is the spin-like topology that a Dirac-like equation can explain in a one-dimensional (1D) harmonic crystal with masses coupled by harmonic springs.[9,10] In addition, it has been demonstrated that periodic topological elastic systems can sustain geometric phases that are "quantized" (such as the berry phase).[11,12] Understanding topological structures through geometric phases is crucial because of the bulk-edge correspondence principle, which predicts localized states at the boundary between crystals of differing topologies, enabling disorder-resistant one-way information transmission.[13]

Despite progress, gaps remain in understanding the geometric and Berry phases in classical counterparts of qubits, such as elastic bits[14] and phase-bits.[15,16] While geometric phases have been proposed in superconducting circuits since their first demonstration of coherent quantum effects,[17–22] and studies have explored topology and geometric phase in topological elastic oscillation, vibration, and waves,[23] similar research in classical two-level systems is lacking. Nevertheless, geometric phases relate closely to the classical idea of moving a vector parallelly across a curved surface. Building on this insight, in this paper, we theoretically demonstrate the controlled accumulation of a geometric or Berry phase in an elastic bit constructed from coupled granules. We previously established the feasibility of creating a classical analog to the qubit using a granular system.[14] Here, we manipulate the elastic bit using the amplitudes, phases, and frequencies of external drivers and observe the resulting phase accumulation.

Granular systems are notable for their highly adaptable dynamics, ranging from strongly to weakly nonlinear or even linear.[24] The current study calculates the Berry phase for an elastic bit in a harmonically driven linearized granular system, capable of generating coherent superpositions of classical in-phase and out-of-phase states, analogous to quantum spin states.[14] These states are projected onto a Bloch sphere, a geometrical representation of a system's state defined by polar and azimuthal angles.[25] The theoretical derivation of the Berry phase is based on the theory of the Foucault pendulum, where the system's state revolves around a constant axis, representing the adiabatic change.[26] While this paper focuses on calculating the Berry phase in a linearized granular system, future work will explore weakly nonlinear and essentially nonlinear systems. The current linearized granular system's primary assumptions include simplifying nonlinear interactions, uniformity assumptions for damping and mass, homogeneous material properties, and specific boundary settings. These simplifications facilitate the analytical tractability of the model. Such assumptions, however, restrict applicability to small amplitude oscillations, excluding significant nonlinear behaviors that may occur in real-world settings. To address these limitations and improve generalizability, future studies will explore nonlinear dynamics and diverse operational conditions, thus broadening the applicability and enhancing the robustness of the findings.

The article is organized as follows to sequentially unveil the various facets of this study: Sec. II introduces the granular system as a two-level elastic bit, analogous to a qubit, and illustrates its states on a Bloch sphere. Section III covers the application and manipulation of these states using quantum-analog logical gates, underscoring their significance in quantum computing analog. Section IV delves into the Berry phase calculations, explaining its role in understanding the system's vibrational behavior. Section V explores the topology of the granular system through Berry phase values, examining its capability to generate a range of states and the impact of external drivers on topological transitions. Finally, Sec. VI draws the conclusions.

## II. REPRESENTATION OF A TWO-LEVEL ELASTIC BIT ON A BLOCH SPHERE

This section presents a detailed study of a nonlinear granular system, linearized and represented as a two-level elastic bit analogous to a qubit, with its states depicted on a Bloch sphere. For such, we study a granular system consisting of two elastically coupled granules driven harmonically [Fig. 1(a)].

The system is governed by the following equations:

$$\begin{aligned} m\ddot{u}_1 &= k_{NL}\left[F_1 e^{i\omega_D t} - u_1 + \sigma_0\right]_+^{3/2} - k_{NL}\left[u_1 - u_2 + \sigma_0\right]_+^{3/2} - \eta\dot{u}_1, \\ m\ddot{u}_2 &= -k_{NL}\left[u_2 - F_2 e^{i\omega_D t} + \sigma_0\right]_+^{3/2} + k_{NL}\left[u_1 - u_2 + \sigma_0\right]_+^{3/2} - \eta\dot{u}_2. \end{aligned} \tag{1}$$

Here, $u_1$ and $u_2$ represent the displacements of the granules from equilibrium, $F_1$ and $F_2$ are the driving amplitudes, and $\omega_D$ is the driving frequency. The term $k_{NL}\left(= \frac{E\sqrt{2R}}{3(1-v^2)}\right)$ denotes the nonlinear stiffness between granules due to Hertzian contact and is dependent on Young's modulus $E$ and Poisson's ratio $v$ of the granules. $\eta$ is the damping coefficient, and $m\left(= \frac{4}{3}\pi\rho R^3\right)$ is the granule mass, which depends on the radius $R$ and density $\rho$ of the granules. The static overlap, $\sigma_0$, simulates pre-compression in the system.

Higher pre-compression reduces the dynamic displacement amplitude, allowing us to approximate the system as linear. Linearization is achieved by expanding the nonlinear term in Eq. (1) using a Taylor series and retaining only the linear component. The resulting linearized equations are[27]

$$\begin{aligned} m\ddot{u}_1 &= k_L\left(F_1 e^{i\omega_D t} - u_1\right) - k_L(u_1 - u_2) - \eta\dot{u}_1, \\ m\ddot{u}_2 &= -k_L\left(u_2 - F_2 e^{i\omega_D t}\right) + k_L(u_1 - u_2) - \eta\dot{u}_2, \end{aligned} \tag{2}$$

where $k_L = \frac{3}{2}k_{NL}\sigma_0^{\frac{1}{2}}$ is the linear coupling stiffness.

All parameters employed in this study are non-dimensionalized to facilitate generalized analysis and to focus on the intrinsic dynamics of the system without the constraints of specific physical units.

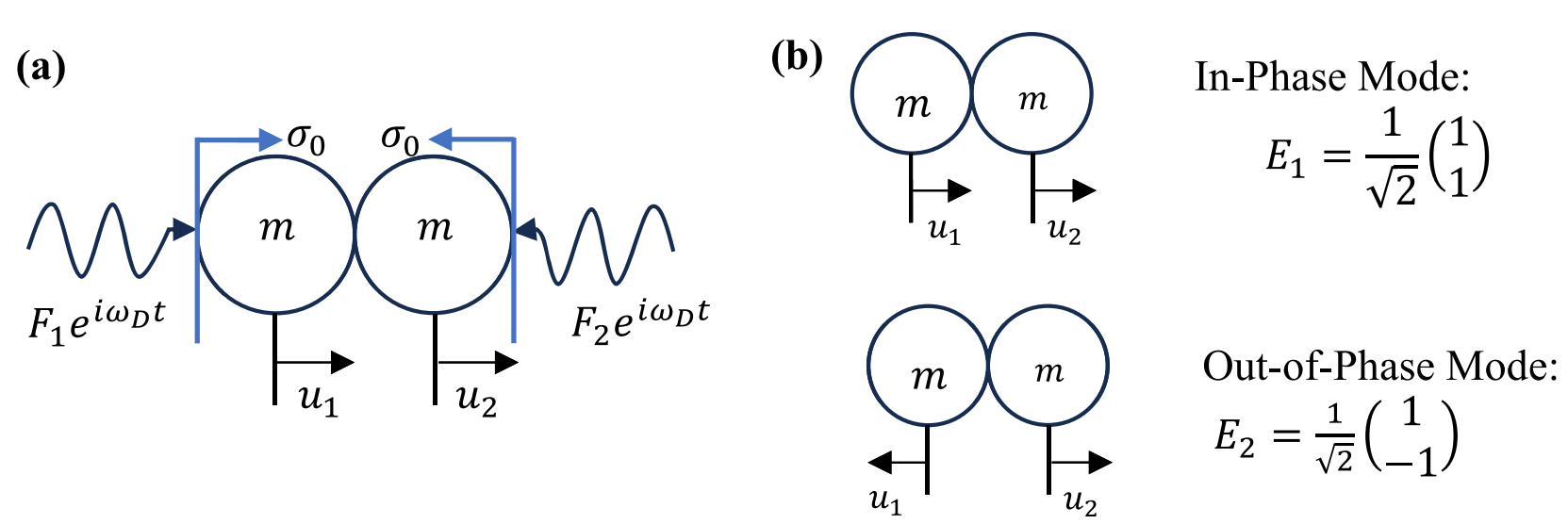

**FIG. 1.** (a) The schematic illustration of the granular system features two elastically coupled granules. These granules are subjected to two external harmonic excitations ($F_1e^{i\omega_D t}$ and $F_2e^{i\omega_D t}$), and a static pre-compression ($\sigma_0$) is applied to control the system's nonlinearity. (b) The two eigenstates of the linearized granular system are represented by the in-phase mode, which corresponds to the $E_1 = \frac{1}{\sqrt{2}}\begin{pmatrix}1\\1\end{pmatrix}$ state, and the out-of-phase mode of the granules, which corresponds to the $E_2 = \frac{1}{\sqrt{2}}\begin{pmatrix}1\\-1\end{pmatrix}$ state.

Furthermore, the damping coefficient $\eta$ is intentionally chosen to be small to ensure the system remains underdamped. This choice is pivotal for maintaining sustained oscillations in response to external excitations as an overdamped system would fail to exhibit the oscillatory behavior essential for demonstrating the accumulation of the Berry phase.

In our non-dimensional framework, we have chosen a damping coefficient of 0.003, which represents minimal energy dissipation. This allows the system to oscillate with negligible attenuation over the timescales considered. This small damping ensures that the steady-state response is predominantly governed by external driving forces rather than internal dissipation. Consequently, even though a mass with a unit velocity experiences only 0.003 N of resistance in our non-dimensional terms, this effectively models a scenario where damping is present but does not overshadow the driving dynamics. By selecting a small $\eta$, we preserve the system's sensitivity to external parameter variations, thereby enabling the exploration of topological transitions and Berry phase accumulation without significant distortion from damping effects.

Equation (2) characterizes a linearized granular system subjected to external driving. In solving this ordinary differential equation (ODE), the general solution comprises both a homogeneous and a particular solution. The homogeneous solution, which arises from the system's intrinsic dynamics, typically involves terms of the form $e^{i\omega t}$. However, due to the presence of damping (as indicated by the damping coefficient $\eta$), the homogeneous solution inherently decays over time, becoming negligible as $t$ approaches infinity.

In practical scenarios, especially when analyzing steady-state behavior, the transient effects captured by the homogeneous solution dissipate, leaving the particular solution as the dominant component. The particular solution directly corresponds to the external driving force and naturally adopts the form $e^{i\omega_D t}$, aligning with the driving frequency $\omega_D$. This approach simplifies the analysis by focusing on the sustained oscillations induced by the external excitation, which are of primary interest in our study.

Therefore, by assuming a solution of the form $e^{i\omega_D t}$, we effectively capture the system's steady-state response to the external driving, bypassing the transient dynamics that fade due to damping. This assumption is well-justified and aligns with standard practices in analyzing driven oscillatory systems.

Assuming solutions of the form $u_1 = A_1 e^{i\omega_D t}$ and $u_2 = A_2 e^{i\omega_D t}$, where $A_1$ and $A_2$ are the amplitude of the vibration displacement, we identify the eigenmodes $E_1 = \frac{1}{\sqrt{2}}\begin{pmatrix}1\\1\end{pmatrix}$ and $E_2 = \frac{1}{\sqrt{2}}\begin{pmatrix}1\\-1\end{pmatrix}$, corresponding to in-phase and out-of-phase modes of the linearized granular system (2), respectively [Fig. 1(b)]. These modes form an orthogonal basis for the system, allowing us to express the displacement field as a linear superposition of $E_1$ and $E_2$,[14]

$$\vec{U} = \begin{pmatrix}u_1\\u_2\end{pmatrix} \equiv \left(\alpha|E_1\rangle + \beta|E_1\rangle\right)e^{i\omega_D t}. \tag{3}$$

Here, $\alpha$ and $\beta$ are the coefficients for $E_1$ and $E_2$, where it is normalized to $|\alpha|^2 + |\beta|^2 = 1$. In Eq. (3), we use the Dirac notation, an analogy with a quantum system, for vectors and apply it to the elastic states of the system by writing vectors in state space.[14] This is because the vectors $E_1$ and $E_2$ are two mutually orthogonal eigenstates of the system and form an orthonormal basis for a 2D Hilbert space. Furthermore, as seen in Eq. (3), the complex coefficients ($\alpha$ and $\beta$) are dependent on each other through phase and form a coherent superposition of states in the space of two possible forms of vibration ($E_1$ and $E_2$). On that basis, the modal contribution in the mode superposition of the total displacement field can be written in the form of a column displacement state vector, $|\psi\rangle$,

$$|\psi\rangle = \begin{pmatrix}\alpha\\\beta\end{pmatrix}.$$

This framework allows us to create a two-level subsystem, an elastic bit, analogous to a qubit in classical terms, through external drivers.[14] An elastic bit is, therefore, a classical analog with respect to the superposition of a qubit, the critical component of quantum computing platforms. In the current study, we like to depict the states of the elastic bit geometrically using the Bloch sphere. Essentially, each elastic bit will be a vector on the Bloch sphere. This vector will be defined in two coordinates, $\theta$ and $\varphi$, where $\theta$ is the polar angle and $\varphi$ is the azimuthal angle. Therefore, the Bloch sphere will represent a linear combination of $E_1$ and $E_2$ states with complex coefficients that will depend on $\theta$ and $\varphi$, indicated by an arrow. We can describe the state space of the granular system using the parameters of the Bloch sphere as follows (see the Appendix for details):

$$|\psi\rangle = \begin{pmatrix}\cos\frac{\theta}{2}\\ e^{i\varphi}\sin\frac{\theta}{2}\end{pmatrix}, \tag{4}$$

where the expression of polar ($\theta$) and azimuthal ($\varphi$) angles are

$$\begin{aligned}\theta &= 2\cos^{-1}(|\alpha|) = 2\cos^{-1}\left[\frac{(F_1+F_2)k_L}{\sqrt{2\left[\left(-m\omega_D^2+k_L\right)^2+\eta^2\omega_D^2\right]}}\right],\\ \varphi &= \arg(\alpha) - \arg(\beta) = \arg\left(\frac{k_L(F_1+F_2)}{\sqrt{2}\left(-m\omega_D^2+k_L+i\eta\omega_D\right)}\right)\\ &\quad - \arg\left(\frac{k_L(F_1-F_2)}{\sqrt{2}\left(-m\omega_D^2+3k_L+i\eta\omega_D\right)}\right).\end{aligned} \tag{5}$$

The current study focuses on the Berry phase and the topology of the linearized granular system as described in Eq. (1) and represented in Eqs. (4) and (5). To investigate these aspects, we apply an external harmonic excitation, driving both granules with differing force amplitudes. The driving force takes the form

$$F_1 = \mathcal{E} + (1-\mathcal{E})e^{i\delta}, \quad F_2 = \mathcal{E} - (1-\mathcal{E})e^{i\delta}. \tag{6}$$

Equation (6) illustrates that the system is driven in such a manner that the amplitudes of the external forces depend on the ratio of two pure states ($\mathcal{E}$) of the linearized granular system and their phase difference $\delta$. In experiments, these parameters can be adjusted using external signal generators. We will show that with the addition of the drivers' frequency together with the drivers' amplitudes $F_1$ and $F_2$, various superpositions of the $E_1$ and $E_2$ states can be achieved. This will allow us to examine their respective state vectors on the Bloch sphere. Consequently, we will demonstrate that even though system (2) is homogeneous, adjusting the external drivers' amplitudes and frequencies leads to different Berry phases and, therefore, different topologies. This approach contrasts with previous studies that altered the system's topology by changing internal parameters, such as the system's mass or the stiffness of its couplings. Our prior work showed that changing internal parameters, such as the stiffness of springs, could generate a Berry phase.[11,12]

To investigate the behavior of the elastic bit on the Bloch sphere, we adjust the parameters $\mathcal{E}$ and $\delta$ as described in Eq. (6). Figure 2 shows how these changes affect the polar angle ($\theta$) and azimuthal angle ($\varphi$). We can visually trace the evolution of these angles on the Bloch sphere. This visualization demonstrates how varying $\mathcal{E}$ and $\delta$ influences the superposition states of the elastic bit. Specifically, it reveals the range and nature of superposition states achievable through parameter manipulation. The Bloch sphere thus serves as a useful tool for comprehending how different parameters alter the system's states.

To investigate scenarios where the azimuthal angle $\varphi$ remains constant while the polar angle $\theta$ varies, we keep $\delta$ fixed and adjust the ratio of $\mathcal{E}$. At $\theta = 0$, a pure state of $E_1$ is achieved, and at $\theta = \pi$, a pure state of $E_2$ is obtained [see Fig. 2(a)]. As $\theta$ changes from 0 to $\pi$ [Fig. 2(a)], it signifies the generation of any superposition of states, incorporating both pure states $E_1$ and $E_2$. The superpositions $\left(\frac{|E_1\rangle+|E_2\rangle}{\sqrt{2}}\right)$ and $\left(\frac{|E_1\rangle+i|E_2\rangle}{\sqrt{2}}\right)$ are realized when the initial phase $\delta$ is 0 and $\pi/2$, respectively. In addition, the superposition of $\left(\frac{|E_1\rangle-|E_2\rangle}{\sqrt{2}}\right)$ is observed in the case of $\delta = \pi$.

We then explore another scenario where $\theta$ is fixed and $\varphi$ varies, achieved by different parameter adjustments. As demonstrated in Fig. 2(b), by setting $\mathcal{E}$ to 0.5 and varying $\delta$ from $-\pi$ to $+\pi$, we observe a constant $\theta$ while $\varphi$ completes a full $2\pi$ rotation. This observation is crucial for calculating the Berry phase, which requires a constant $\theta$ and a varying $\varphi$ that completes a $2\pi$ loop, a topic we will elaborate on in Sec. IV. Altering $\mathcal{E}$ to a different value changes $\theta$. By adjusting $\delta$ from $-\pi$ to $+\pi$ for this specific $\theta$, we can derive another Berry phase value, as we will discuss in Sec. IV. This demonstrates our ability to obtain any combinations of $\theta$ and $\varphi$ values, hence the ability to navigate the Bloch sphere using external drivers.

## III. REALIZATION OF QUANTUM ANALOG GATES WITH AN ELASTIC BIT

Moving forward from the detailed representation of a two-level elastic bit on a Bloch sphere in Secs. II and III focuses on the practical application and manipulation of these representations. Revisiting Fig. 2(a), we noticed that the pure states $E_1$ and $E_2$ manifest under different driving parameters. Equation (4) reveals that these pure states occur at polar angles $\theta$ of 0 and $\pi$, respectively, while the azimuthal angle $\varphi$ indicates the phase difference between $E_1$ and $E_2$. Furthermore, the equal superpositions of states, $\left(\frac{|E_1\rangle+|E_2\rangle}{\sqrt{2}}\right)$ and $\left(\frac{|E_1\rangle+i|E_2\rangle}{\sqrt{2}}\right)$, are represented on the Bloch sphere at coordinates $(\theta, \varphi) = (\pi/2, 0)$ and $(\pi/2, \pi/2)$, and the opposite superpositions of these states are represented by $\left(\frac{|E_1\rangle-|E_2\rangle}{\sqrt{2}}\right)$ and $\left(\frac{|E_1\rangle-i|E_2\rangle}{\sqrt{2}}\right)$ located at $(\theta, \varphi) = (\pi/2, 0)$ and $(\pi/2, 2\pi/3)$. These superposed states are achieved through combinations of $\mathcal{E}$ and $\delta$, induced by external excitations. This section explains how these states change and demonstrates the application of several quantum-analogous logical gates, illustrating the convertibility to superposed states and their manipulation on the Bloch sphere using an elastic bit. The logic gates are essential to quantum-analog computing, similar to classical logic gates in standard computers. They are unitary operators that allow for state changes, crucial for the function of quantum-inspired algorithms, by utilizing superposition and entanglement principles.[28] Here, we demonstrate the quantum analogous gate in the elastic bit and their corresponding unitary transformation matrix, which has a basis state of $E_2 = \frac{1}{\sqrt{2}}\begin{pmatrix}1\\1\end{pmatrix}$ and $E_2 = \frac{1}{\sqrt{2}}\begin{pmatrix}1\\-1\end{pmatrix}$.

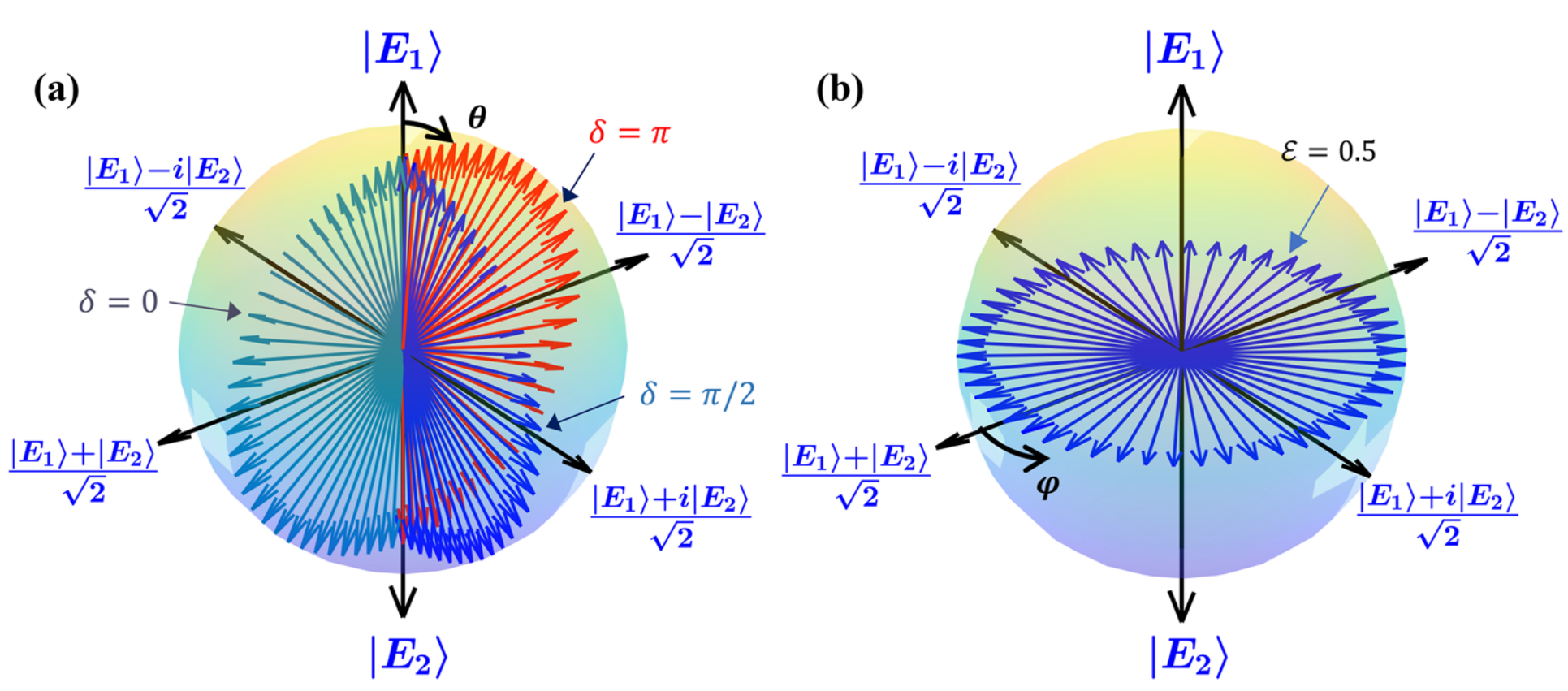

**FIG. 2.** Illustration of the transformation of states of the two-level elastic bit system, represented geometrically on the Bloch sphere. Panel (a) shows the variation in the polar angle ($\theta$) with the ratio of $\mathcal{E}$ while keeping the azimuthal angle ($\varphi$) constant, illustrating the transition from pure states $E_1$ ($\theta = 0$) to $E_2$ ($\theta = \pi$) and various superpositions in between. The initial phase $\delta$ was fixed during this demonstration. Panel (b) focuses on the azimuthal angle ($\varphi$) while keeping $\theta$ constant, by varying $\delta$ from $-\pi$ to $+\pi$ for a set value of $\mathcal{E} = 0.5$. This panel effectively visualizes the full $2\pi$ rotation of $\varphi$ on the Bloch sphere, crucial for the calculation of the Berry phase. The arrows on the sphere indicate the direction of state vector evolution for different parameter adjustments, thereby providing a clear visual guide to understanding how external parameters such as $\mathcal{E}$ and $\delta$ influence the states of the elastic bit. System parameters: $m = 1$, $k_L = 1$, $\eta = 0.003$, $\omega_D = \sqrt{2}$.

In Fig. 3, we illustrate the transformations of elastic bit states through quantum-analog gate operations as follows: (i) from one pure state to another [Figs. 3(a) and 3(b)], (ii) from a pure state to a superposition of states, or vice versa [Fig. 3(c)], and (iii) from one superposition of states to another [Figs. 3(d)–3(f)]. Specifically, the Pauli X gate [Fig. 3(a)] converts the state $|E_1\rangle$ to $|E_2\rangle$ and vice versa, representing a bit flip in the quantum analog system. In contrast, the Pauli Y gate [Fig. 3(b)] transforms $|E_1\rangle$ to $i|E_2\rangle$ and $|E_2\rangle$ to $-i|E_1\rangle$, and vice versa. This indicates that the Pauli X and Y gates alter the state $|E_1\rangle$ to $|E_2\rangle$ along different trajectories, signifying distinct operational paths. The corresponding transformation matrices for these state changes are

$$\begin{bmatrix}1 & 0\\ 0 & -1\end{bmatrix}|E_1\rangle = |E_2\rangle \text{ and } \begin{bmatrix}1 & 0\\ 0 & -1\end{bmatrix}|E_2\rangle = |E_1\rangle,$$

$$\begin{bmatrix}0 & i\\ -i & 0\end{bmatrix}|E_1\rangle = i|E_2\rangle \text{ and } \begin{bmatrix}0 & i\\ -i & 0\end{bmatrix}|E_2\rangle = -i|E_1\rangle.$$

The unitary operation, represented by the matrix $\begin{bmatrix}1 & 0\\ 0 & -1\end{bmatrix}$, serves as the quantum-analog counterpart to the classical NOT gate, commonly known as the Pauli X gate in quantum computing. Unlike the transformation matrix $\begin{bmatrix}0 & 1\\ 1 & 0\end{bmatrix}$ used in quantum mechanics, our analogous system employs a different matrix. This change arises because the computational bases in a true quantum system are $|0\rangle = \begin{pmatrix}1\\0\end{pmatrix}$ and $|1\rangle = \begin{pmatrix}0\\1\end{pmatrix}$. However, in our elastic bit system, we have defined the bases as $|E_1\rangle = \frac{1}{\sqrt{2}}\begin{pmatrix}1\\1\end{pmatrix}$ and $|E_2\rangle = \frac{1}{\sqrt{2}}\begin{pmatrix}1\\-1\end{pmatrix}$. Furthermore, within the $|E_1\rangle$ and $|E_2\rangle$ bases, the Pauli Y matrix is denoted by $\begin{bmatrix}0 & i\\-i & 0\end{bmatrix}$. The Hadamard gate, as illustrated in Fig. 3(c), transforms the pure states. Specifically, it maps $|E_1\rangle$ to $\left(\frac{|E_1\rangle+|E_2\rangle}{\sqrt{2}}\right)$ and $|E_2\rangle$ to $\left(\frac{|E_1\rangle-|E_2}{\sqrt{2}}\right)$. This transformation signifies the mapping of a pure state to a linear combination of superposed states, and vice versa, through the Hadamard matrix $\frac{1}{\sqrt{2}}\begin{bmatrix}1 & 1\\1 & -1\end{bmatrix}$ as follows:

$$\frac{1}{\sqrt{2}}\begin{bmatrix}1 & 1\\1 & -1\end{bmatrix}|E_1\rangle = \left(\frac{|E_1\rangle + |E_2\rangle}{\sqrt{2}}\right); \frac{1}{\sqrt{2}}\begin{bmatrix}1 & 1\\1 & -1\end{bmatrix}|E_2\rangle = \left(\frac{|E_1\rangle - |E_2\rangle}{\sqrt{2}}\right).$$

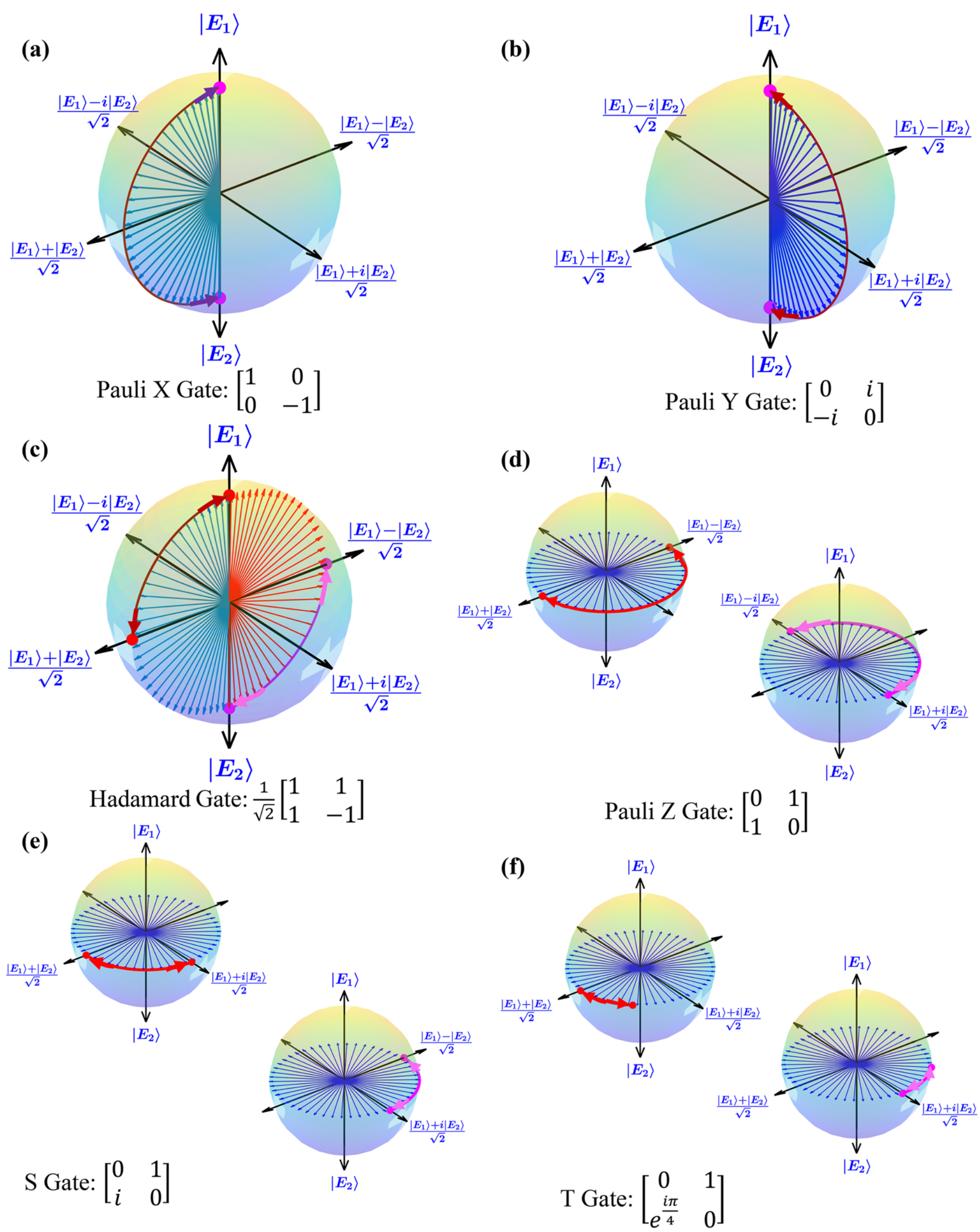

**FIG. 3.** Transformation of the elastic bit states through various quantum-analog logical gates. Each panel represents a different type of gate operation: panels (a) and (b) display the effects of the Pauli X and Y gates, respectively, showcasing state conversions ($|E_1\rangle$ to $|E_2\rangle$ and vice versa) along distinct operational paths. Panel (c) depicts the Hadamard gate's role in mapping pure states to equal superpositions and conversely. Panels (d)–(f) focus on the phase shift gate families, Pauli Z, S, and T gates, respectively. These gates demonstrate the transformation of superposed states by altering only one aspect of the complex amplitude's coefficient, thus modifying the state's direction through the azimuthal angle $\varphi$. The specific transformations affected by each gate are represented through transitions on the Bloch sphere, with arrows indicating the direction and nature of state changes. System parameters: $m = 1, k_L = 1, \eta = 0.003, \omega_D = \sqrt{2}$.

The Pauli Z, S, and T gates [demonstrated in Figs. 3(d)–3(f)] are instrumental in transitioning from one superposed state to another. These gates modify only one aspect of the complex amplitude's coefficient: they alter the state's direction through the azimuthal angle $\varphi$ by $\pi$, $\pi/2$, and $\pi/4$, respectively. The Pauli Z gate, often referred to as a phase flip, uniquely affects the linear equal superposition of states. While it leaves $|E_1\rangle$ and $|E_2\rangle$ states unchanged, it transforms the state $\left(\frac{|E_1\rangle+|E_2\rangle}{\sqrt{2}}\right)$ into $\left(\frac{|E_1\rangle-|E_2\rangle}{\sqrt{2}}\right)$ and $\left(\frac{|E_1\rangle+i|E_2\rangle}{\sqrt{2}}\right)$ to $\left(\frac{|E_1\rangle-i|E_2\rangle}{\sqrt{2}}\right)$ and vice versa, as can be seen through the following transformation matrix $\begin{bmatrix}0 & 1\\1 & 0\end{bmatrix}$ [refer to Fig. 3(d)]:

$$\begin{bmatrix}0 & 1\\1 & 0\end{bmatrix}\left(\frac{|E_1\rangle + |E_2\rangle}{\sqrt{2}}\right) = \left(\frac{|E_1\rangle - |E_2\rangle}{\sqrt{2}}\right),$$

$$\begin{bmatrix}0 & 1\\1 & 0\end{bmatrix}\left(\frac{|E_1\rangle + i|E_2\rangle}{\sqrt{2}}\right) = \left(\frac{|E_1\rangle - i|E_2\rangle}{\sqrt{2}}\right).$$

The other set of gates, as shown in Figs. 3(e) and 3(f), belong to the phase shift gate family. These gates function by mapping the superposition of states. Specifically, they map the state $\left(\frac{|E_1\rangle+|E_2\rangle}{\sqrt{2}}\right)$ to $e^{i\varrho}\left(\frac{|E_1\rangle+|E_2\rangle}{\sqrt{2}}\right)$ and the state $\left(\frac{|E_1\rangle+i|E_2\rangle}{\sqrt{2}}\right)$ to $e^{i\varrho}\left(\frac{|E_1\rangle+i|E_2\rangle}{\sqrt{2}}\right)$, where $\varrho$ represents the phase shift. The same goes for their opposite counterparts. For the S gate [Fig. 3(e)], $\varrho$ is equal to $\pi/2$, and for the T gate [Fig. 3(f)], it is equal to $\pi/4$. The operations of these gates are as follows:

$$\begin{bmatrix}0 & 1\\i & 0\end{bmatrix}\left(\frac{|E_1\rangle + |E_2\rangle}{\sqrt{2}}\right) = e^{\frac{i\pi}{2}}\left(\frac{|E_1\rangle + |E_2\rangle}{\sqrt{2}}\right);$$

$$\begin{bmatrix}0 & 1\\i & 0\end{bmatrix}\left(\frac{|E_1\rangle + i|E_2\rangle}{\sqrt{2}}\right) = e^{\frac{i\pi}{2}}\left(\frac{|E_1\rangle + i|E_2\rangle}{\sqrt{2}}\right).$$

$$\begin{bmatrix}0 & 1\\e^{\frac{i\pi}{4}} & 0\end{bmatrix}\left(\frac{|E_1\rangle + |E_2\rangle}{\sqrt{2}}\right) = e^{\frac{i\pi}{4}}\left(\frac{|E_1\rangle + |E_2\rangle}{\sqrt{2}}\right);$$

$$\begin{bmatrix}0 & 1\\e^{\frac{i\pi}{4}} & 0\end{bmatrix}\left(\frac{|E_1\rangle + i|E_2\rangle}{\sqrt{2}}\right) = e^{\frac{i\pi}{4}}\left(\frac{|E_1\rangle + i|E_2\rangle}{\sqrt{2}}\right).$$

In summary, in this section, we delve into the manipulation of elastic bit states using quantum-analog gate operations, which are mathematically represented by transformation matrices. To provide a clearer physical understanding of how external drivers facilitate these transformations, we elaborate on the interplay between the external parameters and the resulting state changes.

The transformation matrices are inherently linked to the external drivers' characteristics—specifically their frequencies ($\omega_D$) and amplitudes ($F_1$ and $F_2$). These drivers impose specific excitation conditions on the granular system, effectively dictating how the system transitions between different elastic states. By carefully tuning the amplitudes and frequencies of the external drivers, we can engineer the desired state transformations analogous to quantum gate operations. For instance, let us consider the Pauli X gate operation, which flips the state from $|E_1\rangle$ to $|E_2\rangle$ and vice versa. This transformation is achieved by setting the external drivers' amplitudes and frequencies such that the system's response mirrors the matrix multiplication corresponding to the Pauli X gate. Similarly, other gates, such as the Hadamard or Pauli Y gate, are realized by configuring the drivers to induce superpositions or phase shifts in the elastic bit states. To bridge the mathematical formalism with the physical implementation, we have introduced detailed explanations and illustrative examples in Sec. II. Equations (4) and (5) demonstrate how the state space parameters on the Bloch sphere—specifically the polar ($\theta$) and azimuthal ($\varphi$) angles—are directly influenced by the external drivers' frequencies and amplitudes. In addition, Eq. (6) explicates how the phase difference ($\delta$) between the drivers can be manipulated to achieve precise control over the state superpositions. Hence, by varying $\omega_D$, $F_1$, and $F_2$ correspond to different rotation angles and axes on the Bloch sphere, effectively mapping to quantum gate operations such as the Pauli X, Y, and Z gates, as well as the Hadamard and phase shift gates. For instance, adjusting the phase difference $\delta$ between $F_1$ and $F_2$ allows us to implement phase shifts, while altering the amplitude ratio $\mathcal{E}$ facilitates state flips analogous to the Pauli X and Y gates. Finally, we incorporate illustrative examples (Fig. 3), demonstrating how specific combinations of external driver parameters result in the application of these transformation matrices. By linking the physical adjustments of the external drivers to the mathematical form of the gate operations, we provide an intuitive understanding of how quantum-analog logic gates are realized within our classical elastic bit system.

## IV. BERRY PHASE OF THE ELASTIC BIT

Building on the foundations laid in Secs. II and III, where we explored the representation of a two-level elastic bit on a Bloch sphere and its practical application in quantum-analog gates, in this section, we delve into the intricate calculations of the Berry phase for the elastic bit. This exploration is crucial as it characterizes the vibrational behavior and the topological nature of the granular system.

The traditional conception of the Berry phase involves an eigenvector tracing a closed path within the momentum space of periodic systems, such as 1D crystals.[29] This geometric phase arises due to the adiabatic and cyclic evolution of the system's parameters, typically in the context of quantum mechanical systems where the momentum plays a pivotal role in defining the manifold.

In our study, we extend the concept of the Berry phase to a classical mechanical system—the two-level elastic bit—where the manifold is not inherently tied to periodic structures or momentum space. Instead, the manifold is constructed from the external drivers' parameters, specifically the amplitudes and phase differences of the driving forces. This approach does not contradict the traditional understanding but rather generalizes the Berry phase concept to broader contexts, including classical and non-periodic systems.

To elucidate, we define the Berry phase within our system as the geometric phase acquired by the elastic bit's state vector as it undergoes a closed loop in the parameter space formed by the external drivers. This parameter space encompasses variables such as the ratio of excitation amplitudes ($\mathcal{E}$) and the phase difference ($\delta$) between the drivers. By varying these external parameters cyclically, the state vector traces a closed path on the Bloch sphere, analogous to how eigenvectors trace paths in momentum space in quantum systems.

To formulate the mathematical expressions for the Berry connection and Berry phase, we employ a specific ansatz discussed in Sec. II. Previous studies indicate that different ansatzes yield consistent results.[11,12] We will demonstrate that the trajectory of the elastic bit's state vector in parameter space, when mapped onto the Bloch sphere, can be precisely controlled using the external drivers' amplitude, phase, and frequency. This manipulation results in a specific Berry phase. This approach differs from previous methods, where the Berry connection is defined by the variation in complex amplitude unit vectors $A_1$ and $A_2$ in a space parameterized by wave number. In our study, we discretize the polar and azimuthal angles $\theta$ and $\varphi$ and base the Berry connection on these values,[29]

$$\overrightarrow{BC}(A_1, A_2) = A_1^* \cdot (A_1 + \Delta A_1) + A_2^* \cdot (A_2 + \Delta A_2). \tag{7}$$

Equation (7) describes the Berry connection for a two-mass system, extendable to systems with more masses. For example, using three coupled granules creates an elastic trit, an elastic analog of a qutrit.[30] Our focus is on representing the states of the elastic bit on a Bloch sphere, defining the Berry connection in relation to $\theta$ and $\varphi$ as follows:

$$\begin{aligned}\overrightarrow{BC}(\theta,\varphi) = \frac{1}{2}\Bigg[&\left(\cos\frac{\theta}{2} + e^{i\varphi}\sin\frac{\theta}{2}\right)^* \cdot \Bigg(\cos\frac{(\theta+\Delta\theta)}{2} \\ &+ e^{i(\varphi+\Delta\varphi)}\sin\frac{(\theta+\Delta\theta)}{2}\Bigg) + \left(\cos\frac{\theta}{2} - e^{i\varphi}\sin\frac{\theta}{2}\right)^* \\ &\cdot \left(\cos\frac{(\theta+\Delta\theta)}{2} - e^{i(\varphi+\Delta\varphi)}\sin\frac{(\theta+\Delta\theta)}{2}\right)\Bigg].\end{aligned} \tag{8}$$

We use Eq. (8) to calculate the Berry connection followed by the Berry phase associated with polar and azimuthal angles. The Berry phase signifies the total phase acquired by the unit vector tracing a closed path in a manifold (see Fig. 2). This manifold is formed by the evolution of elastic bit states influenced by $\mathcal{E}$ and $\delta$, reflecting the topological characteristics of the system. In cyclic adiabatic processes, the Berry phase emerges as the phase difference over one complete cycle, determined by the geometric properties of the parameter space in the Hamiltonian.[31] For the elastic bit, we can calculate this phase difference using either the amplitudes of the granules [Eq. (7)] or the Bloch sphere angles [Eq. (8)]. Our analysis centers on the cyclic evolution of the Berry curvature in relation to changes in the polar and azimuthal angles as $i\langle\psi|\frac{d}{d(\theta,\varphi)}|\psi\rangle$.

For the case of adiabatic transformation of the system, we take the derivative of the state concerning $\theta$ and $\varphi$ and evolve around the state. Hence, we get $\langle\psi|\frac{d}{d\theta}|\psi\rangle = \frac{1}{2}\begin{bmatrix}\cos\frac{\theta}{2} & e^{-i\varphi}\sin\frac{\theta}{2}\end{bmatrix}\begin{bmatrix}-\frac{1}{2}\sin\frac{\theta}{2} \\ \frac{1}{2}e^{i\varphi}\cos\frac{\theta}{2}\end{bmatrix} = 0$ and $\langle\psi|\frac{d}{d\varphi}|\psi\rangle = \frac{1}{2}\begin{bmatrix}\cos\frac{\theta}{2} & e^{-i\varphi}\sin\frac{\theta}{2}\end{bmatrix}\begin{bmatrix}0 \\ ie^{i\varphi}\sin\frac{\theta}{2}\end{bmatrix} = i\sin^2\frac{\theta}{2}$. For such, the berry phase becomes

$$\begin{aligned}\gamma &= i\int_{\theta_i}^{\theta_f}\int_{\varphi_i}^{\varphi_f}\langle\psi|\frac{d}{d(\theta,\varphi)}|\psi\rangle d(\theta,\varphi) \\ &= i\int_{\theta_i}^{\theta_f}\langle\psi|\frac{d}{d\theta}|\psi\rangle d\theta + i\int_{\varphi_i}^{\varphi_f}\langle\psi|\frac{d}{d\varphi}|\psi\rangle d\varphi = i\int_{\varphi_i}^{\varphi_f} i\sin^2\frac{\theta}{2}d\varphi \\ &= -\frac{1}{2}(\varphi_f - \varphi_i)(1 - \cos\theta).\end{aligned} \tag{9}$$

The above-mentioned expression of the berry phase requires the polar angle $\theta$ to be constant, and the azimuthal angle $\varphi$ must complete a total revolution to close the loop. As shown in Fig. 2(b), keeping the polar angle $\theta$ constant and completing a full revolution of $\varphi$ allow us to close the loop.

Figure 4 demonstrates how Berry phase values vary between 0 and $\pi$ in response to changes in the amplitudes and frequencies of external drivers. A complete $2\pi$ revolution of $\varphi$ is necessary to determine a specific Berry phase. Since the calculation of the Berry phase is conditioned by the existence of a closed path in the parameter space, a closed path revolution of $2\pi$ radians of the azimuthal angle $\varphi$ is necessary to determine a specific Berry phase. This condition ensures that the elastic bit unit vector traces a closed path on the manifold, where the manifold is formed by the evolution of elastic bit states (see Fig. 2). Moreover, as previously demonstrated, elastic bit states are influenced by two parameters: $\mathcal{E}$, which is the ratio of two pure states of the linearized granular system, and $\delta$, which represents their phase difference. To achieve the required $2\pi$ revolution, we vary and discretize the phase $\delta$ from $-\pi$ to $+\pi$ while keeping the state ratio $\mathcal{E}$ constant. This process guarantees a periodic change in the state, resulting in a full $2\pi$ revolution of the azimuthal angle $\varphi$ with a constant polar angle $\theta$, thereby facilitating the computation of the Berry phase. This approach maintains the topological essence of the Berry phase as it relies on the geometric properties of the parameter space rather than the specific nature of the underlying physical system.

Focusing on the 3D representation in Fig. 4, we start with a driving frequency of $\omega_D = \sqrt{2}$, which aligns with the frequencies used in Figs. 2 and 3. At $\omega_D = \sqrt{2}$, the Berry phase attains a trivial value of 0 when the ratio $\mathcal{E}$ is either 0 or 1. Notably, an $\mathcal{E}$ ratio of 1 corresponds to the in-phase vibration mode, and a ratio of 0 corresponds to the out-of-phase mode of the linearized granular system. Conversely, when $E_1$ and $E_2$ are in equal superposition ($\mathcal{E} = 0.5$), the Berry phase reaches a nontrivial value of $\pi$. Such findings highlight the elastic bit system's versatility in producing trivial and nontrivial phases, typically associated with integral multiples of $\pi$. It is important to note that the occurrence of a nontrivial Berry phase of $\pi$ depends not only on $\mathcal{E}$ but also on the driving frequency $\omega_D$, as depicted in Fig. 4. Altering the driving frequency from $\omega_D = \sqrt{2}$ allows tuning $\mathcal{E}$ to reach the same nontrivial Berry phase of $\pi$. Thus, Fig. 4 depicts how the frequencies and amplitudes of the external drivers' interplay to achieve a nontrivial Berry phase. In Fig. 4, the range of $\omega_D$ is limited between 1 and $\sqrt{3}$, corresponding to the eigenfrequencies of the linearized granular system, thereby enabling the system to vibrate in either pure in-phase, out-of-phase, or superposed states. Further analysis of Fig. 4 reveals that a driving frequency equal to the in-phase eigenfrequency ($\omega_D = 1$) necessitates an $\mathcal{E}$ ratio of 0 to achieve a nontrivial Berry phase of $\pi$. Conversely, when the driving frequency matches the out-of-phase eigenfrequency ($\omega_D = \sqrt{3}$), an $\mathcal{E}$ ratio of 1 is required for the same nontrivial Berry phase. The following expression can summarize these observations:

$$\mathcal{E} = \left(2\omega_D^2 + \omega_{01}^2 - \omega_{02}^2\right)/4; \quad \omega_{01} < \omega_D < \omega_{02}. \tag{10}$$

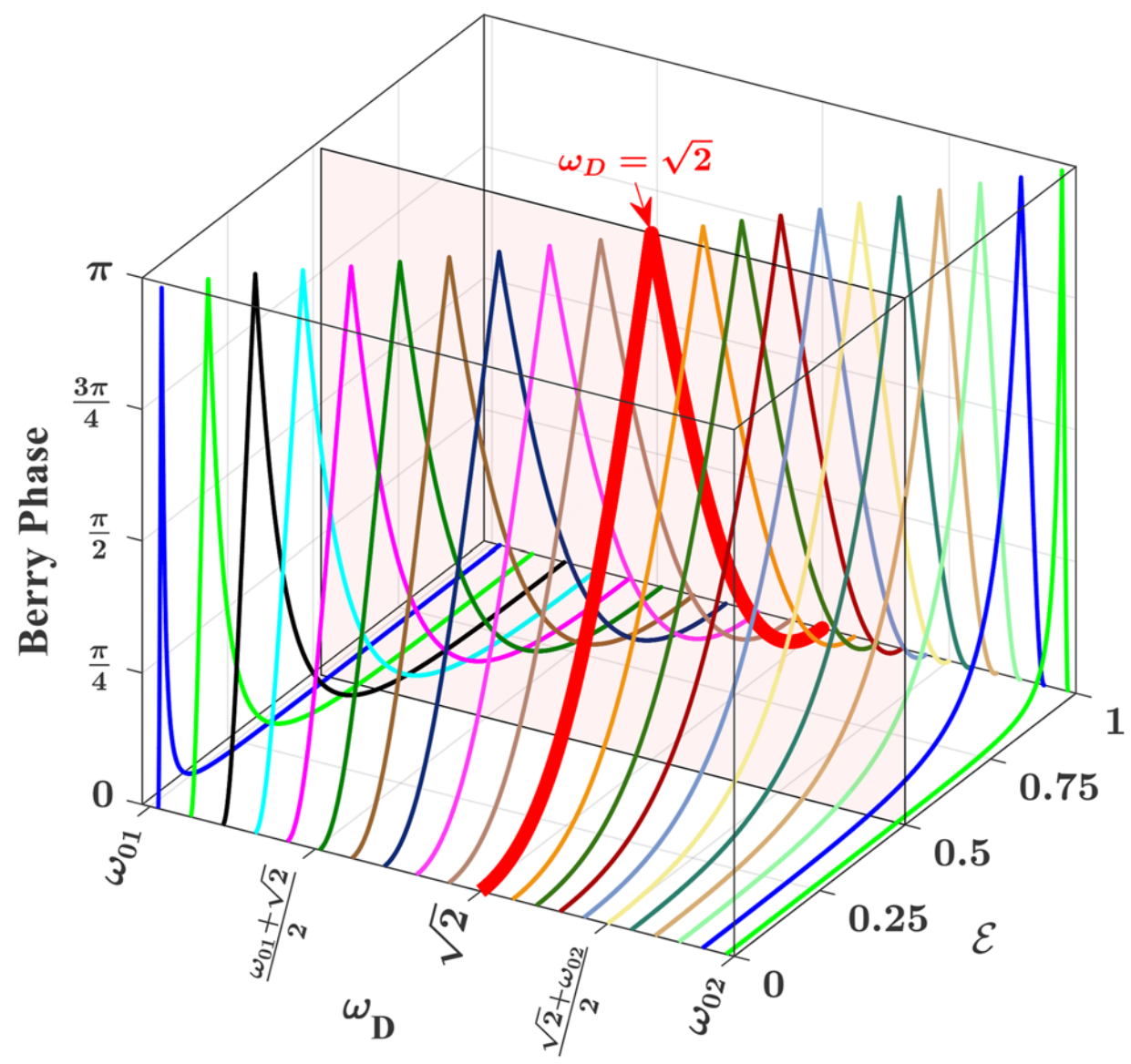

**FIG. 4.** A 3D representation of the Berry phase variation in response to external drivers underscores the system's ability to produce a range of Berry phase values between 0 (trivial phase) and $\pi$ (nontrivial phase). Here, $\omega_{01} = 1$ and $\omega_{02} = \sqrt{3}$ are the two eigenmode frequencies for the linearized granular system.

Here $\omega_{01} = 1$ and $\omega_{02} = \sqrt{3}$ are the two eigenmode frequencies for the coupled granular system. Superpositions other than this specific $\mathcal{E}$ value yield Berry phase values that are not quantized (i.e., neither 0 nor $\pi$). This demonstrates the system's capability to generate a full spectrum of Berry phase values.

## V. SYSTEM DYNAMICS AND TOPOLOGICAL CHARACTERISTICS VIA BERRY PHASE

Through strategic manipulation of external drivers, in the preceding sections, we showcased how distinct Berry phase values emerge, reflecting the system's capability to generate a spectrum of states, ranging from trivial to nontrivial phases. Building on this foundation, this section focuses on understanding the topology of the linearized granular system through Berry phase values since prior research has linked the trivial and nontrivial topologies of systems to integral multiples of $\pi$, including 0.[11,12] Interestingly, although the current system (2) is homogeneous, the Berry phase is not quantized, i.e., is not limited to 0 or $\pi$, contrary to that noted in Ref. 11.

As demonstrated before, Fig. 2 illustrates how the states of the elastic bit evolve, parameterized by the external parameters $\mathcal{E}$ and $\delta$, forming a manifold. The Berry phase represents the total phase gained by the unit vector across the manifold for a closed path. As demonstrated in Fig. 4, setting ($\omega_D$, $\mathcal{E}$) to ($\sqrt{2}$, 0.5) results in a nontrivial Berry phase of $\pi$. Achieving this Berry phase involves cycling $\delta$ from $-\pi$ to $\pi$. If we now focus on the dynamics of the granules with external drivers inputs of ($\omega_D$, $\mathcal{E}$, $\delta$) = ($\sqrt{2}$, 0.5, $\pi$) or ($\omega_D$, $\mathcal{E}$, $\delta$) = ($\sqrt{2}$, 0.5, $-\pi$), we observe that the system oscillates such that only the first mass vibrates, localizing all energy there and resulting in the second mass being silent, as depicted in Fig. 5(a). Figure 5(a) plots each mass's amplitude as $\delta$ cycles from $-\pi$ to $\pi$, with $\omega_D$ and $\mathcal{E}$ set at $\sqrt{2}$ and 0.5, respectively. Conversely, inputs of ($\omega_D$, $\mathcal{E}$, $\delta$) = ($\sqrt{2}$, 0.5, 0) reverse the response, directing all energy to the second mass and nullifying the first mass's response [Fig. 5(a)]. Hence, the nontrivial Berry phase signifies a distinct coupled vibration, with energy alternating between granules and oscillation/rest cycles. Alternatively, setting ($\omega_D$, $\mathcal{E}$) to either ($\sqrt{2}$, 0) or ($\sqrt{2}$, 1) yields a trivial Berry phase of 0 (as is shown in Fig. 4), where both masses receive equal energy and maintain identical amplitudes, regardless of $\delta$ [Fig. 5(b)].

The dynamics of the system and their correlation with the Berry phase are further explored by examining the phase differences between the masses (denoted by $\phi_{m_1-m_2}$) and between the masses and the drivers. For simplicity, we focus on the phase difference between the first mass and the first driver (denoted by $\phi_{m_1-d_1}$). When $\mathcal{E}$ is equal to 0 or 1, resulting in a Berry phase of 0, $\phi_{m_1-m_2}$ is either 0 (in-phase) or $\pi$ (out-of-phase), as shown in Fig. 5(c). However, for other values of $\mathcal{E}$, $\phi_{m_1-m_2}$ varies with $\delta$, even when $\mathcal{E}$ is constant. For $\mathcal{E}$ values above 0.5, $\phi_{m_1-m_2}$ starts at 0 (in-phase) at $\delta = -\pi$, returns to 0 at $\delta = 0$, and follows a similar pattern from 0 to $\pi$. If $\mathcal{E}$ is below 0.5, $\phi_{m_1-m_2}$ begins at $\pi$ (out-of-phase), returning to $\pi$ at $\delta = 0$, with a similar pattern for the remaining $\delta$ values from 0 to $\pi$. Interestingly, at $\mathcal{E} = 0.5$, $\phi_{m_1-m_2}$ remains constant at $\pi/2$ for all $\delta$ values, indicating a lack of phase transition, unlike the dual transitions observed at lower and higher $\mathcal{E}$ values. This $\mathcal{E}$ value coincides with the nontrivial Berry phase of $\pi$.

Focusing on $\phi_{m_1-d_1}$, as depicted in Fig. 5(d), we observe that for $\mathcal{E} = 0$ or 1, $\phi_{m_1-d_1}$ remains constant, regardless of $\delta$ values. Similar to $\phi_{m_1-m_2}$, $\phi_{m_1-d_1}$ varies with $\delta$ for fixed $\mathcal{E}$ values. For $\mathcal{E}$ less than 0.5, $\phi_{m_1-d_1}$ starts at zero phase and returns to it for $\delta$ ranging from $-\pi$ to $\pi$. For $\mathcal{E}$ greater than 0.5, the phase starts at $\pi$ and returns to $\pi$. At $\mathcal{E} = 0.5$, $\phi_{m_1-d_1}$ consistently remains at $\pi/2$, mirroring the behavior of $\phi_{m_1-m_2}$. This again marks the absence of a phase transition, contrasting with the dual transitions at lower and higher $\mathcal{E}$ values. The $\mathcal{E}$ value of 0.5 represents a topological transition point. Below and above this value, $\phi_{m_1-m_2}$ and $\phi_{m_1-d_1}$ exhibit different behaviors. This transition indicates a change in the system's topological properties as external parameters vary. To visualize this, we define two zones: the first corresponding to $\mathcal{E}$ values below 0.5 and the second above 0.5. In Fig. 5(c), the first zone appears upper in the plot, while in Fig. 5(d), it is lower. Reaching the topological transition point at $\mathcal{E} = 0.5$, where the nontrivial Berry phase of $\pi$ is obtained, the zones invert their positions in the plots [Figs. 5(c) and 5(d)]. This inversion results from changes in the drivers' parameters. Unlike previous studies where internal system parameters were used to achieve topological transitions, our study employs external driver parameters for this purpose. For example, Deymier *et al.* in Ref. 32 demonstrated band inversion and topological transitions based on the coupling strength of elastic rotators.

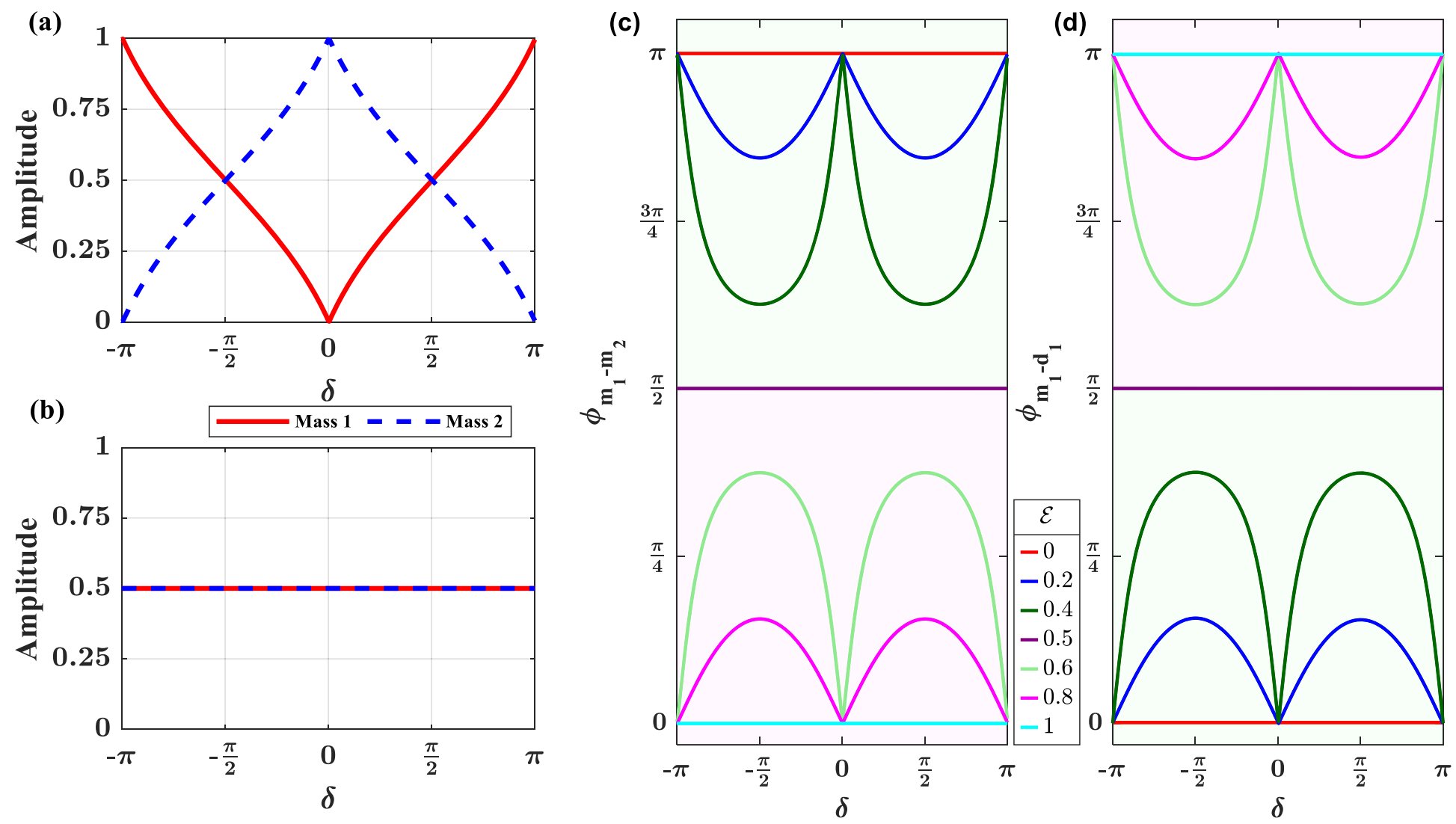

**FIG. 5.** Correlation of the dynamics of the linearized granular system with the calculated Berry phase values, revealing insights into the system's topological nature. *Vibration amplitude responses*: panel (a) shows the amplitude response of each mass as $\delta$ cycles from $-\pi$ to $\pi$, illustrating how energy localization in the granules corresponds to nontrivial Berry phase values; panel (b) contrasts this with scenarios leading to a trivial Berry phase, where both masses exhibit identical amplitude responses. *Phase difference analysis*: panels (c) and (d) explore the phase differences between the granules ($\phi_{m_1-m_2}$) and between the granules and the drivers ($\phi_{m_1-d_1}$). They display how these phase differences vary with $\delta$ for different $\mathcal{E}$ ratios, highlighting the distinct behaviors below and above the $\mathcal{E}$ value of 0.5, which marks a topological transition point.

While this section primarily discusses the characterization of the granular system's dynamics using Berry phase values at the driving frequency $\omega_D = \sqrt{2}$, it is important to note that the same phenomenon occurs at other driving frequencies as well. However, altering the driving frequency shifts the topological transition point to a different value of $\mathcal{E}$, which can be precisely determined from Fig. 4.

## VI. CONCLUDING REMARKS

This study investigated the Berry phase in classical systems, drawing parallels with quantum models, focusing on a two-level elastic bit formed from externally driven coupled granular system. The representation of the elastic bit states on the Bloch sphere has allowed for a clearer visualization and manipulation of these states, a concept previously unexplored in classical systems. This representation is crucial in understanding the dynamics of the system and its transition between various states. We have shown that, similar to quantum systems, classical elastic bits can transform analogous to quantum gates. These transformations, executed through external drivers, enable the manipulation of states on the Bloch sphere, offering insights into the potential of classical systems to mimic quantum operations.

A significant contribution of our research is calculating the Berry phase for an elastic bit. This phase varies depending on the amplitudes, frequencies, and phases of external drivers, highlighting the elastic bit system's capability to generate a broad spectrum of Berry phase values, ranging from trivial to nontrivial. Notably, we identified a nontrivial Berry phase of $\pi$ under certain conditions, associated with a specific coupled vibration mode of the granular system. Therefore, the Berry phase is indicative of the system's vibrational behavior and topological properties. This discovery is pivotal for controlling topological properties in classical systems and exploring the bulk-edge correspondence principle.

The insights from studying the topological characteristics of elastic bits via Berry phase calculations could lead to innovative developments in materials science, particularly in designing materials with distinctive mechanical properties. Extending these concepts to systems with more than two levels, such as elastic trits analogous to qutrits, could further unravel the intricacies of higher-dimensional topological spaces.

## ACKNOWLEDGMENTS

M.A.H. acknowledges partial support from the NSF (Grant Nos. 2204382 and 2242925).

## AUTHOR DECLARATIONS

### Conflict of Interest

The authors have no conflicts to disclose.

### Author Contributions

M.A.H. conceived the idea of the research. All authors analyzed the findings and contributed to the scientific discussion and the manuscript's writing.

**Kazi T. Mahmood**: Conceptualization (equal); Data curation (equal); Formal analysis (equal); Funding acquisition (equal); Investigation (equal); Methodology (equal); Software (lead); Writing – original draft (equal); Writing – review & editing (equal). **M. Arif Hasan**: Conceptualization (lead); Data curation (equal); Formal analysis (equal); Funding acquisition (lead); Investigation (equal); Methodology (equal); Supervision (lead); Writing – original draft (equal); Writing – review & editing (equal).

## DATA AVAILABILITY

The data that support our findings of the present study are available from the corresponding author upon reasonable request.

## APPENDIX: MODAL REPRESENTATION OF BLOCH STATES

We are studying a granular system composed of two elastically coupled granules, formulated using the dynamics equation presented in Eq. (1). Linearization of Eq. (1) is achieved through Taylor series expansion. The expansions for the equations of mass 1 and mass 2 are derived as follows:

$$\begin{aligned} m\ddot{u}_1 &= k_{NL}\left[F_1 e^{i\omega_D t} - u_1 + \sigma_0\right]_+^{\frac{3}{2}} - k_{NL}\left[u_1 - u_2 + \sigma_0\right]_+^{\frac{3}{2}} - \eta\dot{u}_1 \\ \Rightarrow m\ddot{u}_1 &= k_{NL}\sigma_0^{\frac{3}{2}}\left[1 + \frac{F_1 e^{i\omega_D t} - u_1}{\sigma_0}\right]_+^{\frac{3}{2}} - k_{NL}\sigma_0^{\frac{3}{2}}\left(1 + \frac{u_1 - u_2}{\sigma_0}\right)_+^{\frac{3}{2}} - \eta\dot{u}_1 \\ \Rightarrow m\ddot{u}_1 &= k_{NL}\sigma_0^{\frac{3}{2}}\left[1 + \frac{3}{2}\left(\frac{F_1 e^{i\omega_D t} - u_1}{\sigma_0}\right) - \frac{3}{8}\left(\frac{F_1 e^{i\omega_D t} - u_1}{\sigma_0}\right)^2 - \frac{3}{48}\left(\frac{F_1 e^{i\omega_D t} - u_1}{\sigma_0}\right)^3 + \cdots\right]_+ \\ &\quad - k_{NL}\sigma_0^{\frac{3}{2}}\left[1 + \frac{3}{2}\left(\frac{u_1 - u_2}{\sigma_0}\right) - \frac{3}{8}\left(\frac{u_1 - u_2}{\sigma_0}\right)^2 - \frac{3}{48}\left(\frac{u_1 - u_2}{\sigma_0}\right)^3 + \cdots\right]_+ - \eta\dot{u}_1 \\ \Rightarrow m\ddot{u}_1 &= \left[\frac{3}{2}k_{NL}\sigma_0^{\frac{1}{2}}\left(F_1 e^{i\omega_D t} - u_1\right) - \frac{3}{8}k_{NL}\sigma_0^{-\frac{1}{2}}\left(F_1 e^{i\omega_D t} - u_1\right)^2 - \frac{3}{48}k_{NL}\sigma_0^{-\frac{3}{2}}\left(F_1 e^{i\omega_D t} - u_1\right)^3 + \cdots\right]_+ \\ &\quad - \left[\frac{3}{2}k_{NL}\sigma_0^{\frac{1}{2}}(u_1 - u_2) - \frac{3}{8}k_{NL}\sigma_0^{-\frac{1}{2}}(u_1 - u_2)^2 - \frac{3}{48}k_{NL}\sigma_0^{-\frac{3}{2}}(u_1 - u_2)^3 + \cdots\right]_+ - \eta\dot{u}_1 \\ \Rightarrow m\ddot{u}_1 &= \left[k_L\left(F_1 e^{i\omega_D t} - u_1\right) + k_2\left(F_1 e^{i\omega_D t} - u_1\right)^2 + k_3\left(F_1 e^{i\omega_D t} - u_1\right)^3 + \cdots\right]_+ \\ &\quad - \left[k_L(u_1 - u_2) + k_2(u_1 - u_2)^2 + k_3(u_1 - u_2)^3 + \cdots\right]_+ - \eta\dot{u}_1, \end{aligned} \tag{A1}$$

$$\begin{aligned} m\ddot{u}_2 &= -k_{NL}\left[u_2 - F_2 e^{i\omega_D t} + \sigma_0\right]_+^{\frac{3}{2}} + k_{NL}\left[u_1 - u_2 + \sigma_0\right]_+^{\frac{3}{2}} - \eta\dot{u}_2 \\ \Rightarrow m\ddot{u}_2 &= -k_{NL}\sigma_0^{\frac{3}{2}}\left[1 + \frac{u_2 - F_2 e^{i\omega_D t}}{\sigma_0}\right]_+^{\frac{3}{2}} + k_{NL}\sigma_0^{\frac{3}{2}}\left(1 + \frac{u_1 - u_2}{\sigma_0}\right)_+^{\frac{3}{2}} - \eta\dot{u}_2 \\ \Rightarrow m\ddot{u}_2 &= -k_{NL}\sigma_0^{\frac{3}{2}}\left[1 + \frac{3}{2}\left(\frac{u_2 - F_2 e^{i\omega_D t}}{\sigma_0}\right) - \frac{3}{8}\left(\frac{u_2 - F_2 e^{i\omega_D t}}{\sigma_0}\right)^2 - \frac{3}{48}\left(\frac{u_2 - F_2 e^{i\omega_D t}}{\sigma_0}\right)^3 + \cdots\right]_+ \\ &\quad + k_{NL}\sigma_0^{\frac{3}{2}}\left[1 + \frac{3}{2}\left(\frac{u_1 - u_2}{\sigma_0}\right) - \frac{3}{8}\left(\frac{u_1 - u_2}{\sigma_0}\right)^2 - \frac{3}{48}\left(\frac{u_1 - u_2}{\sigma_0}\right)^3 + \cdots\right]_+ - \eta\dot{u}_2 \\ \Rightarrow m\ddot{u}_2 &= -\left[\frac{3}{2}k_{NL}\sigma_0^{\frac{1}{2}}\left(u_2 - F_2 e^{i\omega_D t}\right) - \frac{3}{8}k_{NL}\sigma_0^{-\frac{1}{2}}\left(u_2 - F_2 e^{i\omega_D t}\right)^2 - \frac{3}{48}k_{NL}\sigma_0^{-\frac{3}{2}}\left(u_2 - F_2 e^{i\omega_D t}\right)^3 + \cdots\right]_+ \\ &\quad + \left[\frac{3}{2}k_{NL}\sigma_0^{\frac{1}{2}}(u_1 - u_2) - \frac{3}{8}k_{NL}\sigma_0^{-\frac{1}{2}}(u_1 - u_2)^2 - \frac{3}{48}k_{NL}\sigma_0^{-\frac{3}{2}}(u_1 - u_2)^3 + \cdots\right]_+ - \eta\dot{u}_2 \\ m\ddot{u}_2 &= -\left[k_L\left(u_2 - F_2 e^{i\omega_D t}\right) + k_2\left(u_2 - F_2 e^{i\omega_D t}\right)^2 + k_3\left(u_2 - F_2 e^{i\omega_D t}\right)^3 + \cdots\right]_+ \\ &\quad + \left[k_L(u_1 - u_2) + k_2(u_1 - u_2)^2 + k_3(u_1 - u_2)^3 + \cdots\right]_+ - \eta\dot{u}_2. \end{aligned} \tag{A2}$$

Here, $k_L = \frac{3}{2}k_{NL}\sigma_0^{\frac{1}{2}}$, $k_2 = -\frac{3}{8}k_{NL}\sigma_0^{-\frac{1}{2}}$, and $k_3 = -\frac{3}{48}k_{NL}\sigma_0^{-\frac{3}{2}}$ represent the first, second, and third order coupling stiffness of the granular system, respectively, with $k_{NL}$ being the nonlinear coupling stiffness between the granules. For higher pre-compression ($\sigma_0$), where the relative displacement of the granules is less than pre-compression i.e., if $\sigma_0 \gg u_1 - u_2$, this condition suppresses the higher-order stiffness. Consequently, $k_2$ and $k_3$ are compared to $k_L$. By taking only the linear term from Eqs. (A1) and (A2), we obtain the linearized equation of the coupled granular system as shown in Eq. (2) of the article.

Assuming the solution for the linearized coupled granular network as $u_1 = A_1 e^{i\omega_D t}$ and $u_2 = A_2 e^{i\omega_D t}$ and applying in Eq. (2), we derive the simultaneous equation for the amplitude $A_1$ and $A_2$ as

$$\begin{aligned} -m\omega_D^2 A_1 e^{i\omega_D t} &= k_L\left(F_1 e^{i\omega_D t} - A_1 e^{i\omega_D t}\right) \\ &\quad - k_L\left(A_1 e^{i\omega_D t} - A_2 e^{i\omega_D t}\right) - i\eta\omega_D A_1 e^{i\omega_D t} \\ \Rightarrow -m\omega_D^2 A_1 e^{i\omega_D t} &= \left[k_L(F_1 - A_1) - k_L(A_1 - A_2) - i\eta\omega_D A_1\right]e^{i\omega_D t} \\ \Rightarrow -m\omega_D^2 A_1 &= k_L(F_1 - A_1) - k_L(A_1 - A_2) - i\eta\omega_d A_1 \\ \Rightarrow \left(-m\omega_D^2 + 2k_L + i\eta\omega_D\right)A_1 &- k_L A_2 = k_L F_1, \end{aligned} \tag{A3}$$

$$\begin{aligned} -m\omega_D^2 A_2 e^{i\omega_D t} &= -k_L\left(A_2 e^{i\omega_D t} - F_2 e^{i\omega_D t}\right) \\ &\quad + k_L\left(A_1 e^{i\omega_D t} - A_2 e^{i\omega_D t}\right) - i\eta\omega_D A_2 e^{i\omega_d t} \\ \Rightarrow -m\omega_D^2 A_2 e^{i\omega_D t} &= \left[-k_L(A_2 - F_2) + k_L(A_1 - A_2) - i\eta\omega_D A_2\right]e^{i\omega_D t} \\ \Rightarrow -m\omega_D^2 A_1 &= -k_L(A_2 - F_2) + k_L(A_1 - A_2) - i\eta\omega_D A_2 \\ \Rightarrow -k_L A_1 &+ \left(-m\omega_D^2 + 2k_L + i\eta\omega_D\right)A_2 = k_L F_2. \end{aligned} \tag{A4}$$

By taking the real part of the coefficient of the amplitudes from Eqs. (A3) and (A4), we can write the stiffness matrix of the solution as

$$[K] = \begin{bmatrix} -m\omega_D^2 + 2k_L & -k_L \\ -k_L & -m\omega_D^2 + 2k_L \end{bmatrix}. \tag{A5}$$

The eigenmode frequencies of the system depend on this stiffness matrix. For the eigenmode frequencies, the determinant of the stiffness matrix is det $(K) = 0$. With that, the eigenmode frequencies are

$$\omega_{01} = \sqrt{\frac{k_L}{m}} \quad \omega_{02} = \sqrt{\frac{3k_L}{m}}. \tag{A6}$$

From the solution of the eigenmode frequencies, we observe that the frequencies depend on the mass of the granules and the linearized stiffness between them. Setting $k_L = 1$ and $m = 1$, we find the eigenmode frequencies as $\omega_{01} = 1$ and $\omega_{02} = \sqrt{3}$. By employing the elimination and substitution method on the simultaneous amplitude equations from Eqs. (A3) and (A4), we derive the complex amplitude solutions $A_1$ and $A_2$,

$$\begin{aligned} A_1 &= \frac{F_1 k_L\left(-m\omega_D^2 + 2k_L + i\eta\omega_D\right) + F_2 k_L^2}{\left(-m\omega_D^2 + 2k_L + i\eta\omega_D\right)^2 - k_L^2}, \\ A_2 &= \frac{F_1 k_L^2 + F_2 k_L\left(-m\omega_D^2 + 2k_L + i\eta\omega_D\right)}{\left(-m\omega_D^2 + 2k_L + i\eta\omega_D\right)^2 - k_L^2}. \end{aligned} \tag{A7}$$

Using the complex amplitude ($A_1$ and $A_2$) and the linearized eigen states of $\left(E_1 = \frac{1}{\sqrt{2}}\begin{pmatrix}1\\1\end{pmatrix} \text{ and } E_2 = \frac{1}{\sqrt{2}}\begin{pmatrix}1\\-1\end{pmatrix}\right)$, we can express the displacement field of the granules through the superposition of the states' coefficients $\alpha$ and $\beta$,[14]

$$\begin{aligned} \vec{U} &= \begin{pmatrix}u_1\\u_2\end{pmatrix} = \begin{pmatrix}A_1\\A_2\end{pmatrix}e^{i\omega_D t} = \frac{1}{\sqrt{|\tilde{\alpha}|^2 + |\tilde{\beta}|^2}}\left(\tilde{\alpha}|E_1\rangle + \tilde{\beta}|E_2\rangle\right)e^{i\omega_D t} \\ &= \left(\alpha|E_1\rangle + \beta|E_2\rangle\right)e^{i\omega_D t} \Rightarrow \begin{pmatrix}A_1\\A_2\end{pmatrix} = \frac{1}{\sqrt{2}}\begin{pmatrix}\alpha + \beta\\ \alpha - \beta\end{pmatrix}. \end{aligned} \tag{A8}$$

Substituting the complex amplitude ($A_1$ and $A_2$), we obtain the superposition of states coefficient $\alpha$ and $\beta$,

$$\begin{aligned} \alpha &= \frac{A_1 + A_2}{\sqrt{2}} = \frac{k_L(F_1 + F_2)}{\sqrt{2}\left(-m\omega_D^2 + k_L + i\eta\omega_D\right)}, \\ \beta &= \frac{A_1 - A_2}{\sqrt{2}} = \frac{k_L(F_1 - F_2)}{\sqrt{2}\left(-m\omega_D^2 + 3k_L + i\eta\omega_D\right)}. \end{aligned} \tag{A9}$$

With the two possible forms of eigenmode vibration ($E_1$ and $E_2$), the displacement field vector $\vec{U}$ of the coupled granules, utilizing the coefficient of the complex amplitude, can be expressed as follows:

$$\vec{U} = \alpha|E_1\rangle + \beta|E_2\rangle \ \Rightarrow \vec{U} = r_\alpha e^{i\xi_\alpha}|E_1\rangle + r_\beta e^{i\xi_\beta}|E_2\rangle, \tag{A10}$$

where $r_\alpha$ and $r_\beta$ are the magnitudes of the complex amplitude coefficients of $\alpha$ and $\beta$, respectively. These states are represented in polar space as $r_\alpha = \cos\theta$ and $r_\beta = \sin\theta$, where $\theta$ is the polar angle, indicating the angle of superposition states with the z-axis of a Bloch sphere, as shown in Fig. 6.[33]

In Fig. 6, the z axis in a Bloch sphere represents the pure states $|E_1\rangle$ and $|E_2\rangle$, referred to as the polar angle. Multiplying both sides of Eq. (A10) by the global phase $e^{-i\xi_\alpha}$ allows us to represent the displacement field as $\vec{U}' = e^{-i\xi_\alpha}\vec{U}$.[34] The state space of the granular system can then be represented accordingly,

$$\vec{U}' = r_\alpha|E_1\rangle + r_\beta e^{i(\xi_\beta-\xi_\alpha)}|E_2\rangle\ \vec{U}' = \cos\theta|E_1\rangle + \sin\theta e^{i\varphi}|E_2\rangle. \quad \text{(A11)}$$

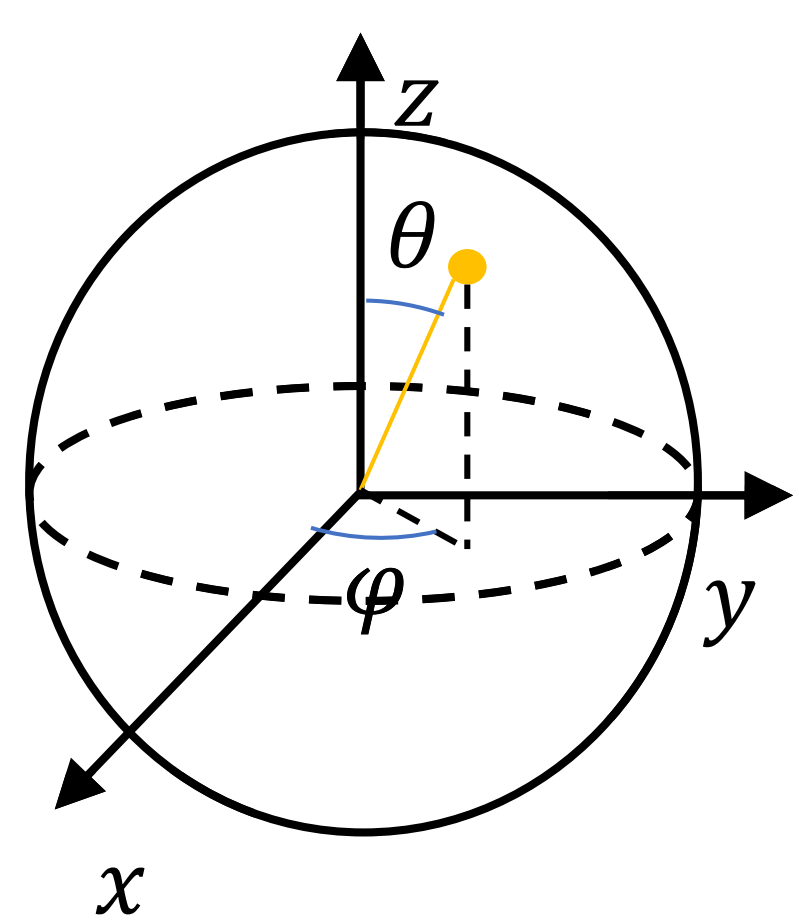

**FIG. 6.** Three-dimensional illustration of the Bloch sphere, showcasing its relationship with the Cartesian coordinate system. This diagram highlights the sphere's axes and the defining angles—theta ($\theta$) and phi ($\varphi$)—representing an elastic bit's state.

Here, the phase difference between the complex amplitude coefficients is referred to as the azimuthal angle $\varphi = \xi_\alpha - \xi_\beta$, which signifies the rotation of the states in the x-y plane as depicted in Fig. 6. To limit the ranges of both angles, we use the half-angle of the polar angle $\theta$ to represent the state on the Bloch sphere. Since the global phase does not alter the superposition of states, we can determine the displacement field as

$$\vec{U} = \cos\frac{\theta}{2}|E_1\rangle + e^{i\varphi}\sin\frac{\theta}{2}|E_2\rangle \text{ where}, 0 \le \theta \le \pi \text{ and } 0 \le \varphi \le 2\pi. \quad \text{(A12)}$$

By the linear combination of $E_1$ and $E_2$ in Eq. (A12), we can express the displacement field vector in terms of Bloch sphere angle $\theta$ and $\varphi$. The modal contribution of the displacement vector $\vec{U}$ is described in a column state vector $|\psi\rangle$ either through the superposition coefficients $\alpha$ and $\beta$ or the Bloch sphere angles $\theta$ and $\varphi$ as stated in Eq. (4),

$$|\psi\rangle = \begin{pmatrix}\alpha\\ \beta\end{pmatrix} = \begin{pmatrix}\cos\frac{\theta}{2}\\ e^{i\varphi}\sin\frac{\theta}{2}\end{pmatrix}.$$